\documentstyle[12pt,epsfig]{article}
\begin{document}

\title{Partial Wave Analysis of $\bar pp \to \bar \Lambda \Lambda$}

\vskip 1mm
\begin {center}
{D.V.~Bugg},   \\
{Queen Mary, University of London, London E1\,4NS, UK}
\vskip 4.5mm
\end {center}

\begin{abstract}
A partial wave analysis of PS185 data for $\bar pp \to \bar \Lambda
\Lambda$ is presented.
A $^3S_1$ cusp is identified in the inverse process
$\bar \Lambda \Lambda \to \bar p p$ at threshold, using
detailed balance to deduce cross sections from
$\bar pp \to \bar \Lambda \Lambda$.
Partial wave amplitudes for $\bar pp$ $^3P_0$, $^3F_3$, $^3D_3$ and
$^3G_3$ exhibit a behaviour very similar to resonances observed in
Crystal Barrel data.
With this identification, the $\bar pp \to \bar \Lambda \Lambda$
data then provide evidence for a new $I = 0$, $J^{PC} = 1^{--}$
resonance with mass $M = 2290 \pm 20$ MeV, $\Gamma = 275 \pm 35$
MeV, coupling to both $^3S_1$ and $^3D_1$.
\end{abstract}

\section {Introduction}
The PS185 collaboration has made extensive measurements of
$\bar pp \to \bar \Lambda \Lambda$ at LEAR.
Integrated cross sections have been measured at fine steps of
momentum close to the $\bar \Lambda \Lambda$ threshold [1-4];
Ref. [4] summarises results.
Differential cross sections extend up to 1990 MeV/c.
The decays of $\Lambda$ and $\bar \Lambda$ analyse their
polarisation $P_y = A_{00N0}=A_{000N}$ and measure spin correlations
$C_{NN}$, $C_{SS}$, $C_{LL}$ and $C_{LS}=C_{SL}$ [5].
Data from a polarised target provide further measurements with
target polarisation normal to the scattering plane [6].

An early partial wave analysis close to threshold was made
by Tabakin et al. [7].
The objective here is to extend the partial wave analysis over the
whole momentum range, including  polarised target data.

There are six spin dependent amplitudes for $\bar pp \to \bar
\Lambda \Lambda $ [8], one more than for $NN$ and $\bar NN$
elastic scattering,  where particles in initial and final
states are identical.
There are 6 further measurements from the polarised
target.
Firstly the asymmetry $A_{0N00}$ from the polarised target is
different to $A_{00N0}$ because the nucleon and $\Lambda$ are different
particles.
Secondly, there are rather precise measurements of spin
transfer parameters $D_{NN}$ and $K_{NN}$.
Thirdly, the triple spin
parameters $C_{0NSS}$, $C_{0NLS}$ and  $C_{0NSL}$ are independent
measurements.
Here, the first suffix refers to the $\bar p$ beam, which
is unpolarised, the second refers to the target proton,
the third refers to the $\bar \Lambda$ and the fourth
refers to the $\Lambda$.
Further measurements of $C_{0NLL}$ and $C_{0NNN}$ are
redundant.
Paschke and Quinn [9] show that $C_{0NSS} = -C_{0NLL}$,
although both sets of data can be included in the analysis, to improve
statistics.
Also $A_{0NNN} = A_{0N00}$; the latter is much better determined
than $A_{0NNN}$.

There is then a chance of determining the six amplitudes up to
an overall unmeasurable phase.
In principle eleven sets of data are sufficient providing they
explore all amplitudes in an ideal way.
In practice, it turns out that the determination is almost unique
at 1637 MeV/c, the only momentum where polarised target data
are available.
There are some minor reservations concerning
relative branchings to $^3F_2$ and $^3P_2$ and between
$^3D_1$ and $^3S_1$.
It is necessary to apply a mild constraint
to partial wave amplitudes for $^3P_2 \to \, ^3F_2$ and $^3F_2 \to
\, ^3P_2$, in order to prevent them drifting to large values.
It is also necessary to make the simplifying assumption
that, away from 1637 MeV/c, $^3D_1 \to \, ^3S_1$ and
$^3D_1 \to \, ^3D_1$ amplitudes are related to $^3S_1 \to \, ^3S_1$
by simple centrifugal barrier factors.
In the limited mass range over which data are available, these are
mild assumptions, which have little effect on the
determination of other partial waves.

The available mass range extends only 200 MeV above the
$\bar \Lambda \Lambda$ threshold.
Resonances typically have widths of 250 MeV, so  it is difficult
to establish the presence of resonances from PS185 data alone.
Nonetheless,  results can be compared with analyses of Crystal
Barrel and PS172 data having the same quantum numbers.
In those data, a mass range of 500 MeV is available.
For $I = 0$, $ C = +1$, there are seven sets of data from these
two experiments for final states $\pi ^0 \pi ^0$,
$\eta \eta$, $\eta \eta '$, $\eta \pi ^0 \pi ^0$, $\eta ' \pi ^0 \pi
^0$, $3\eta$ and $\pi ^- \pi ^+$; in addition there is some information
from the production procss $\bar pp \to (\eta \pi ^0 \pi ^0)\eta$.
From those extensive data, many resonances are observed with securely
determined parameters [10].
It is of interest to see if those resonances corresponds to structures
observed  in $\bar pp \to \bar \Lambda \Lambda$.
That is quite likely, in the  same way that $J^P = 0^+$ resonances
appear in both $\pi \pi $ and  $K\bar K$ channels.

\section {A cusp at the $\bar \Lambda \Lambda$ threshold}
Fig. 1(a) shows integrated cross sections
very close to threshold for $\bar pp \to \bar \Lambda \Lambda$.
The curve shows the S-wave intensity deduced later from
the partial wave analysis; the remaining intensity comes
from P-waves in this mass range.

\begin{figure} [ht]
\begin{center}
\epsfig{file=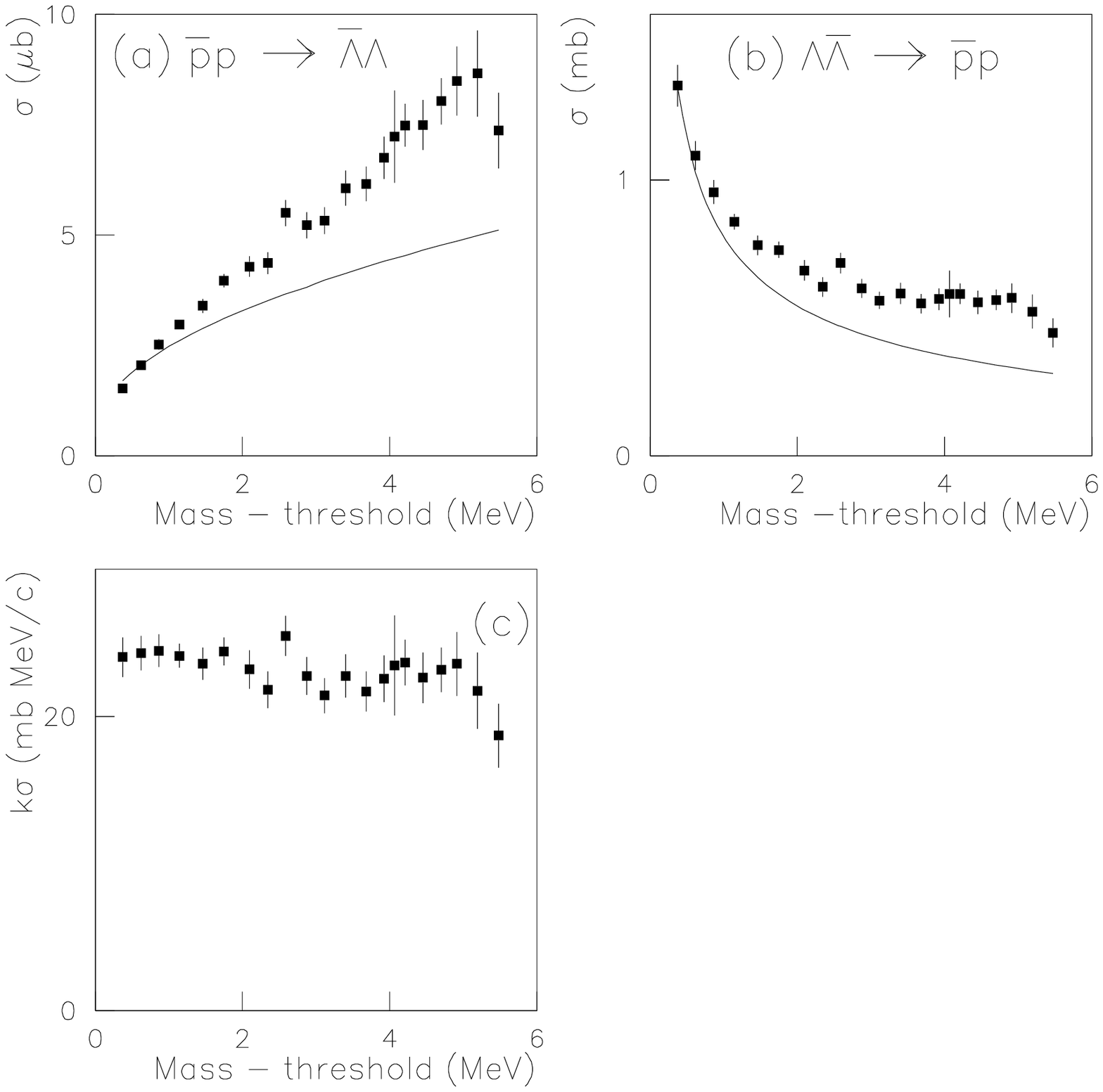,width=12cm}\
\caption{(a) Integrated cross sections for $\bar pp \to \bar
\Lambda \Lambda$; the curve shows the S-wave cross
section from the amplitude analysis; (b) the corresponding
cross section for $\bar \Lambda \Lambda \to \bar pp$;
the curve is the fitted S-wave intensity;
(c) $\sigma (\bar \Lambda \Lambda \to \bar pp ) \times k $ v.
excitation energy, after subtracting the P-wave intensity
.}
\end{center}
\end{figure}

The cross section for the inverse process
$\bar \Lambda \Lambda \to \bar pp$ may be derived using
detailed balance:
\begin {equation}
\sigma (\bar \Lambda \Lambda \to \bar pp ) =
(p/k)^2 \sigma (\bar pp \to \bar \Lambda \Lambda ).
\end {equation}
Here, $p$ and $k$ are momenta of $p$  and $\Lambda$
in the centre of mass frame.
Fig. 1(b) shows the resulting cross sections for
$\bar \Lambda \Lambda \to \bar pp$.
There is a cusp at threshold, first reported in Ref. [11].
Cusps are in principle well known, but are not often seen,
so this case is interesting.

The cusp is a feature of S-waves. The curve shows the
fitted S-wave intensity; in this mass range, the difference
from data is purely due to P-waves.
These P-waves are surprisingly strong near threshold,
but are very well determined from polarisations and
forward-backward asymmetries in differential cross sections
(Fig. 2 below).
The PS185 collaboration makes the reasonable conjecture that
the S-wave is strongly absorbed into other open channels,
whereas in this mass range P-waves are highly peripheral
and therefore suffer little attenuation from annihilation.
Up to 6 MeV, P-waves have momenta $k < 85$ MeV/c, and
therefore a classical impact parameter $> 2.3$ fm.

The curve follows the familiar $1/v$ law of thermal neutron physics.
The $1/v$ dependence is verified in Fig. 1(c), where
$\sigma (\bar \Lambda \Lambda \to \bar pp) \times k$ is plotted
after subtracting off the contributions from P-waves.

It will be useful to exhibit the origin of the cusp assuming
there is an S-wave resonance, which will be fitted later
to the data. The result is however quite general and is derived
in the textbook of Landau and Lifshitz [12].
For an S-wave resonance, the partial wave amplitude is
\begin {equation}
 f_s(\bar pp \to \bar \Lambda \Lambda ) =
\frac {1}{p}\frac {\sqrt {\Gamma _{\bar p p}(s)\Gamma _
{\bar \Lambda \Lambda }(s)}}{D(s)},
\end {equation}
where $D(s) = M^2 - s -m(s) - iM\Gamma _{tot}(s)$;
the term $m(s)$ in the denominator $D(s)$ will be discussed below.
Since $\Gamma _{\bar \Lambda \Lambda }\propto k$ and
$\Gamma _{\bar pp} \propto p$ near threshold,
\begin {equation}
 f_s(\bar pp \to \bar \Lambda \Lambda ) \propto \frac {\sqrt
 {k/p}}{D(s)}.
\end {equation}
The amplitude for $\bar \Lambda \Lambda$ elastic scattering is
\begin {equation}
 f_s(\bar \Lambda \Lambda \to \bar \Lambda \Lambda )
 =\frac {1}{k}\frac {\Gamma _{\bar \Lambda \Lambda }(s)}{D(s)}.
\end {equation}
Apart from a slow energy dependence from $D(s)$, the
amplitude goes to a constant at threshold, namely the
scattering length $a$.
The amplitude for $\bar \Lambda \Lambda \to pp$ is
\begin {equation}
 f_s(\bar \Lambda \Lambda \to \bar pp)
 =\frac {1}{k}\frac {\sqrt {\Gamma _{\bar pp (s)}
 \Gamma _{\bar \Lambda \Lambda }(s)}}{D(s)},
\end {equation}
 and is proportional to $(p/k)^{1/2}/D(s)$ at threshold.
 The intensity $|f_S(\bar \Lambda \Lambda \to \bar pp)|^2 \propto 1/k$,
 apart from the slowly varying factor $p/|D(s)|^2$.
 This is the origin of the $1/v$ law.

 At threshold there is a step in
 $Im \,  f_s(\bar \Lambda \Lambda \to \bar \Lambda \Lambda )$.
 Associated with this step is a rapid variation of
 $Re \,  f_s(\bar \Lambda \Lambda \to \bar \Lambda \Lambda )$,
 i.e. a dispersive effect.
 For a resonance, $m(s)$ of eqn. (3) is given [13] by
\begin {equation}
m(s) = \frac {M^2 - s}{\pi } \int \frac {Im \, f_s (s') \,
ds'}{(M^2 - s')(s' - s)},
\end {equation}
 where a subtraction is made on resonance. This
formula will be used in fitting an S-wave resonance to the data.

 For higher partial waves, the centrifugal barrier makes cusp
 effects negligible.

\section {Data and analysis procedures}
Figs. 2--9 show the PS185 data, together with fits
described here.

\begin{figure} [t]
\begin{center}
\epsfig{file=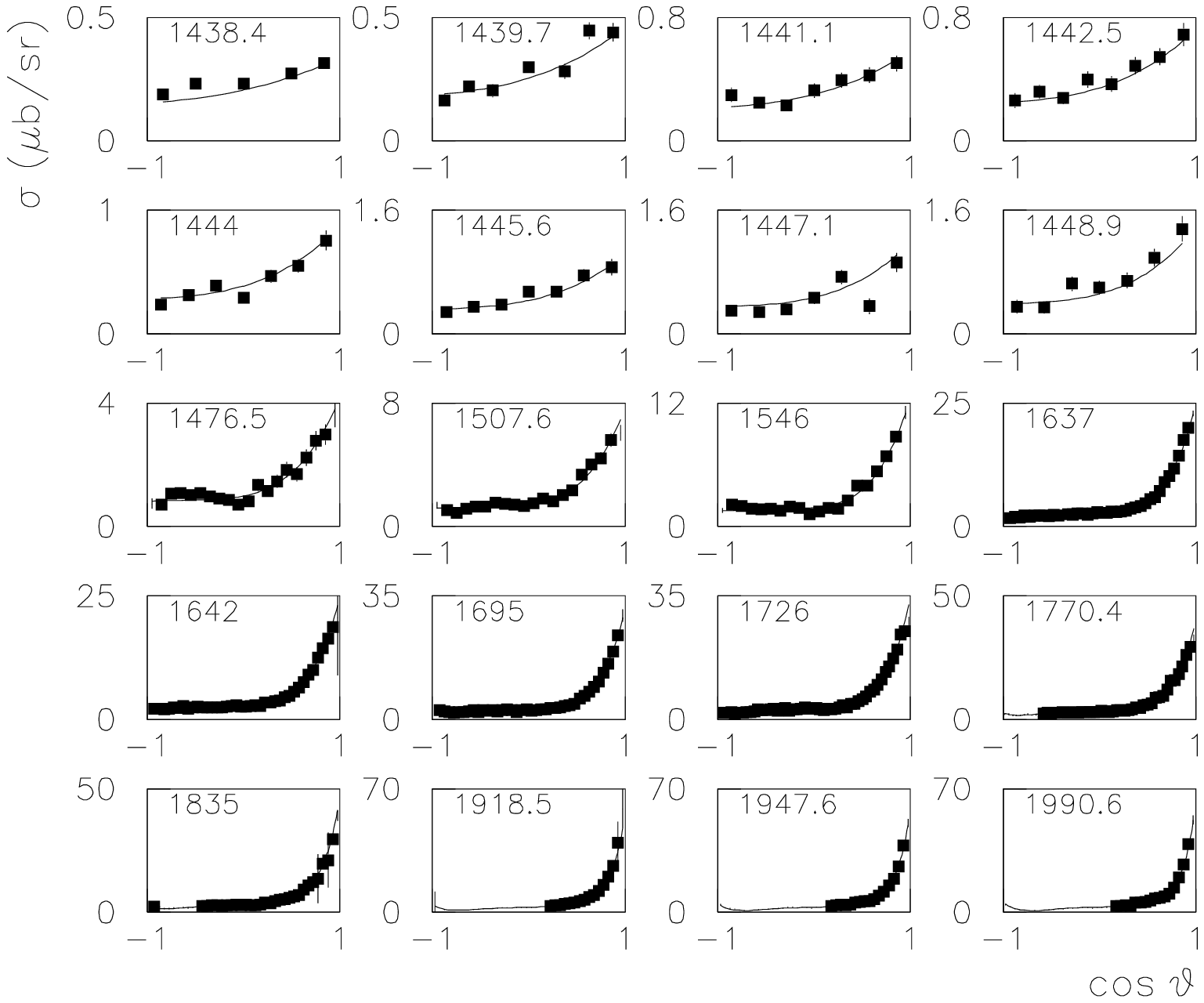,width=16cm}
~\
\caption{Fit to differential cross sections for $\bar pp \to
\bar \Lambda \Lambda$; lab
momenta are indicated in each panel in MeV/c.}
\end{center}
\end{figure}

\begin{figure} [t]
\begin{center}
\epsfig{file=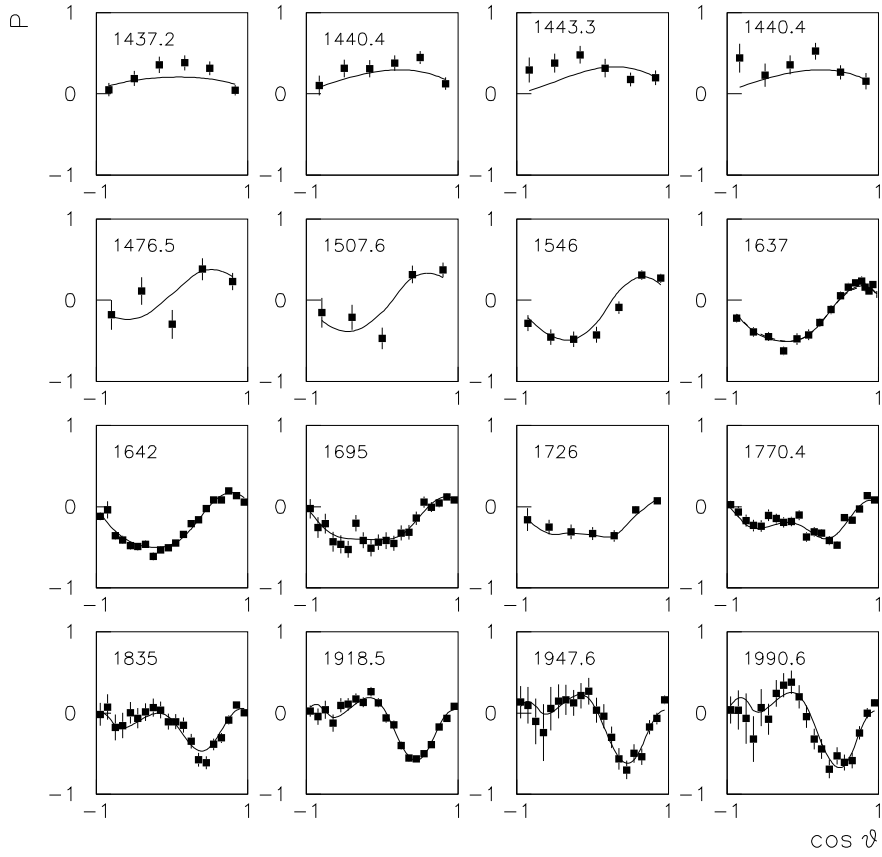,width=13cm}
~\
\caption{Fit to hyperon polarisations $P_y$ for $\bar pp \to
\bar \Lambda \Lambda$.}
\end{center}
\end{figure}

\begin{figure}
\begin{center}
\epsfig{file=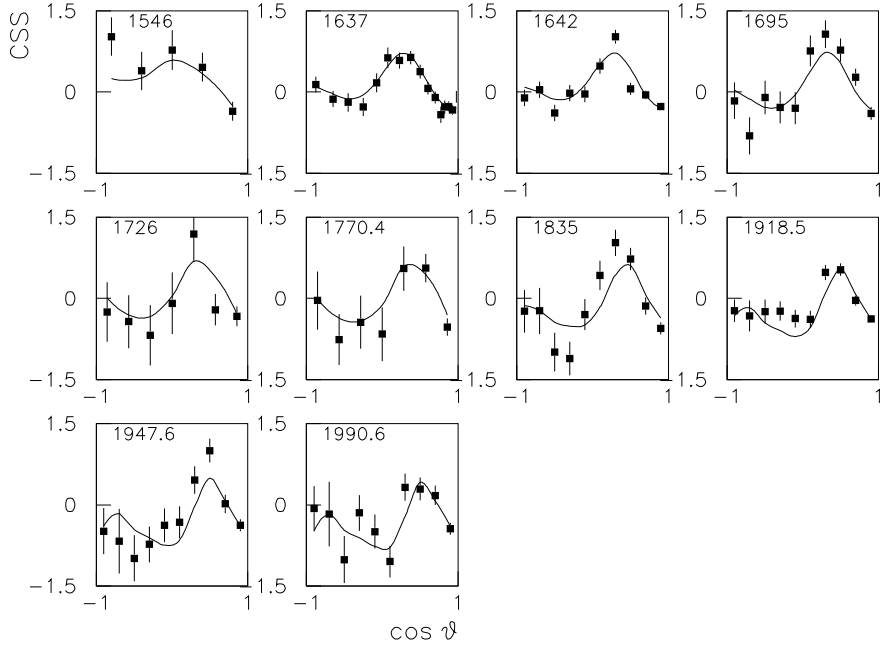,width=13cm}
~\
\caption{Fit to the spin correlation parameter $C_{SS}$
for $\bar pp \to \bar \Lambda \Lambda$;
$S$ is the component of spin transverse to the beam and in the
plane of scattering.}
\end{center}
\end{figure}

\begin{figure}
\begin{center}
\epsfig{file=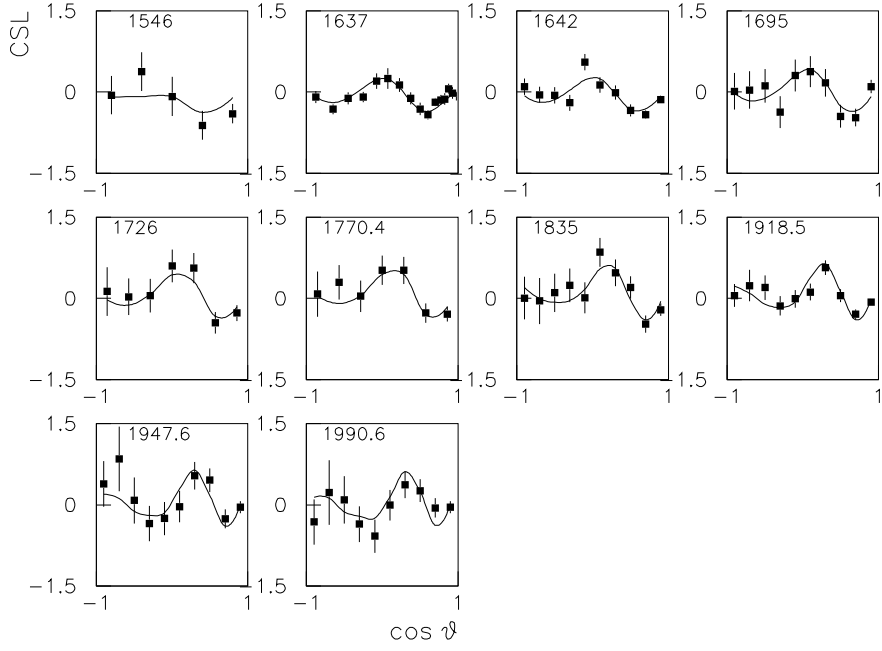,width=13cm}
~\
\caption{Fit to the spin correlation parameter $C_{SL}$ for
$\bar pp \to \bar \Lambda \Lambda$;
$S$ is as in Fig. 4 and $L$ is the longitudinal component of spin.}
\end{center}
\end{figure}

\begin{figure}
\begin{center}
\epsfig{file=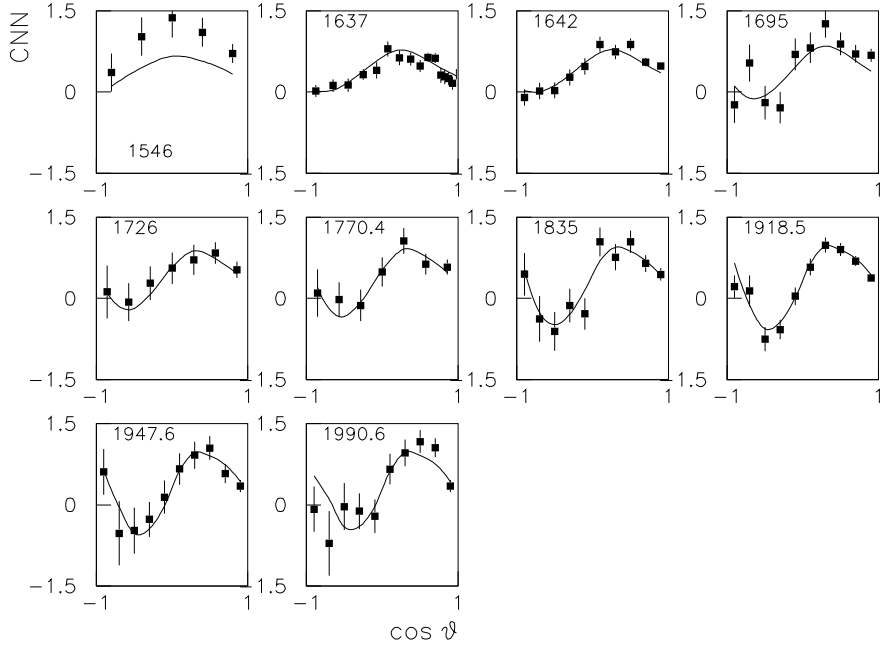,width=13cm}
~\
\caption{Fit to spin correlation parameter $C_{NN}$ for
$\bar pp \to \bar \Lambda \Lambda$; $N$ is the component of
spin normal to the scattering plane.}
\end{center}
\end{figure}

\begin{figure}
\begin{center}
\epsfig{file=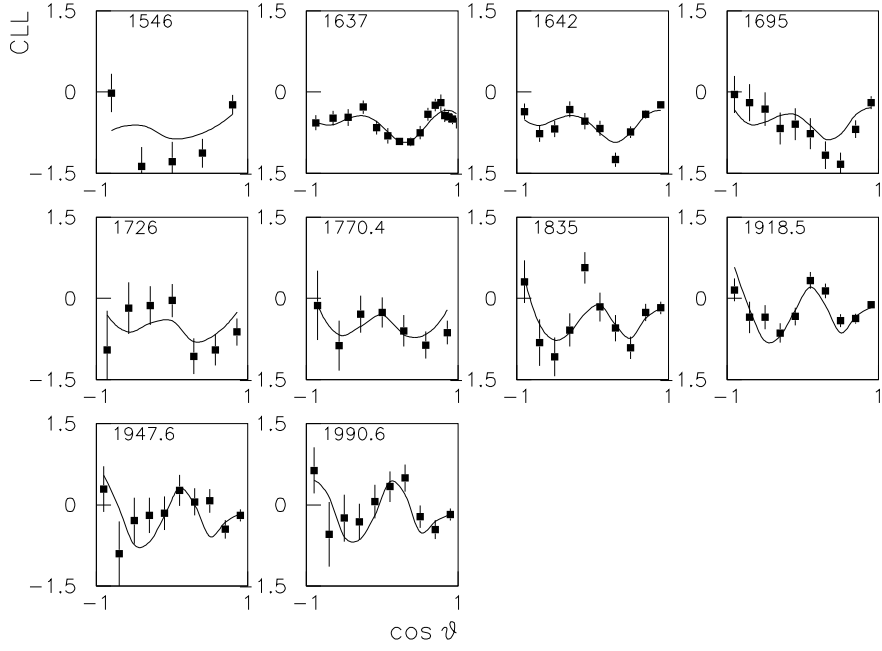,width=13cm}
~\
\caption{Fit to spin correlation parameters $C_{LL}$ for
$\bar pp \to \bar \Lambda \Lambda$; $L$ is the longitudinal
component of spin.}
\end{center}
\end{figure}

\begin{figure}
\begin{center}
\epsfig{file=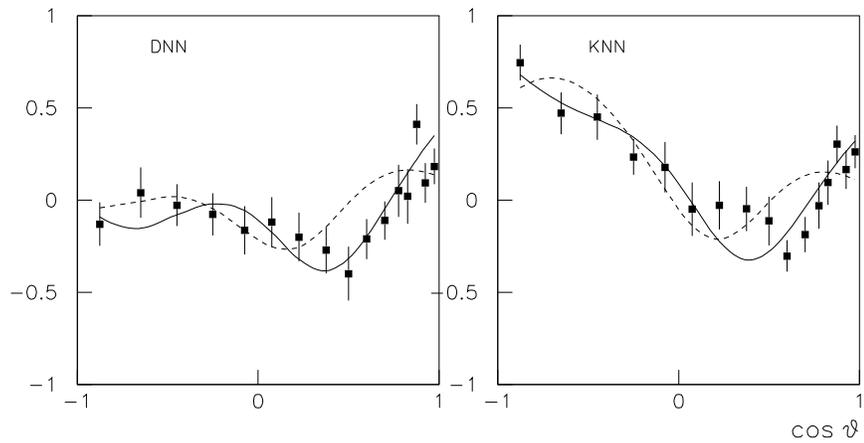,width=13cm}
~\
\caption{Fit to spin transfer parameters for $\bar pp \to \bar
\Lambda \Lambda$; $D$ refers to spin transfer from proton to
$\Lambda$ and $K$ to spin transfer from proton to $\bar \Lambda$;
the dashed curve shows the fit omitting the $^3G_3 \to \, ^3D_3$
amplitude;
data are at a beam momentum of 1637 MeV/c.}
\end{center}
\end{figure}

\begin{figure}
\begin{center}
\epsfig{file=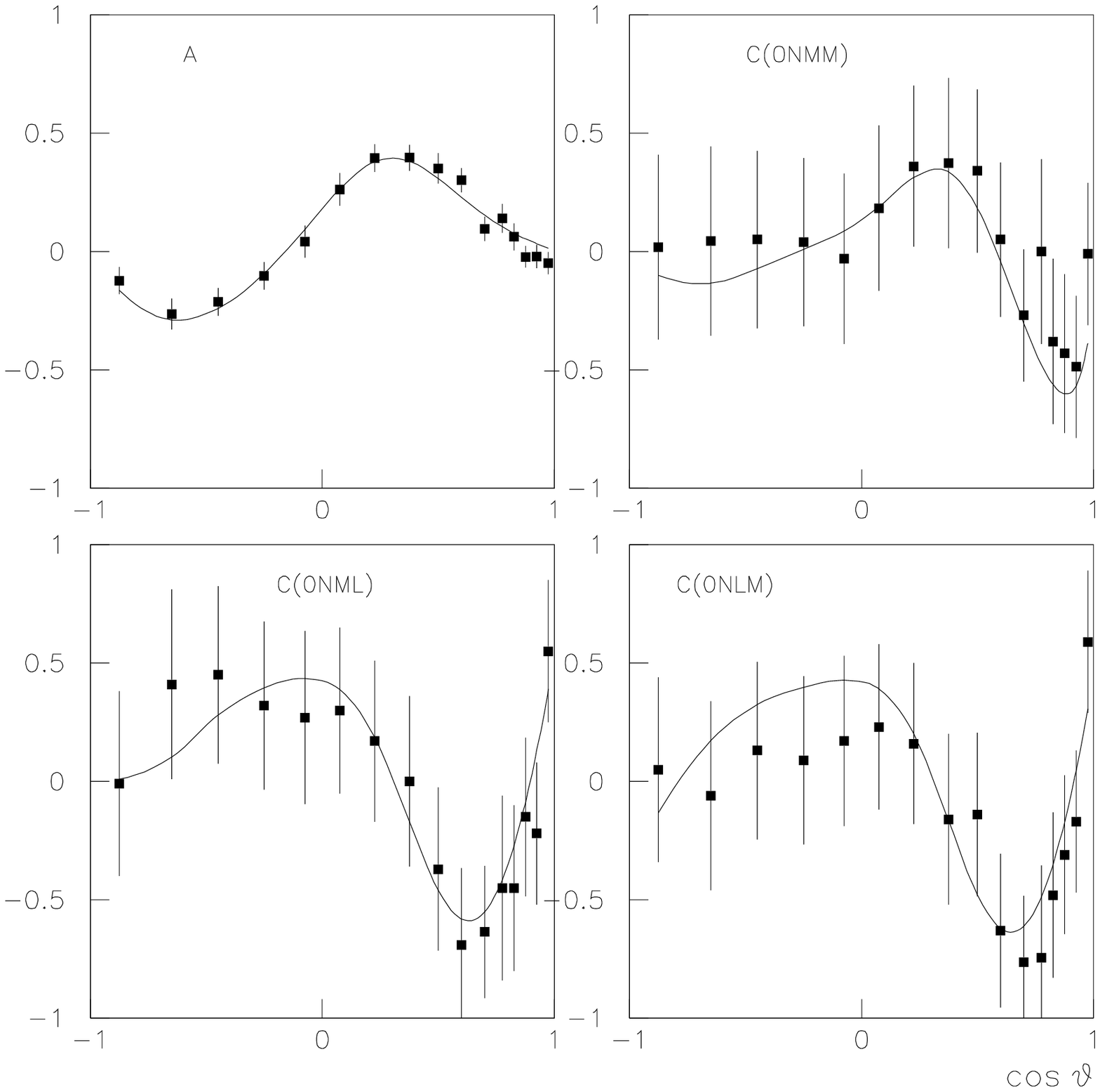,width=13cm}
~\
\caption{Fit to the asymmetry $A$ from the polarised target for
$\bar pp \to \bar \Lambda \Lambda$ and to
triple spin parameters. Data are at a beam momentum of 1637 MeV/c.}
\end{center}
\end{figure}

 \subsection {Formulae for Observables and Partial wave amplitudes}
 Elchikh and Richard [8] show that six amplitudes are needed
 to describe $\bar pp \to \bar \Lambda \Lambda$.
 Formulae for observables are readily adapted from the well known
 expressions for $NN$ elastic scattering [14].
 They have also been written down by Paschke and Quinn [9].
 However, one needs to be aware that Paschke and Quinn quantise
 along the same axes for initial and final states.
 Suppose the $y$-axis is taken normal to the scattering plane,
 $z$ along the beam direction and $x$ sideways in the plane
 of scattering.
 For spin transfer parameters, the expressions of
 Paschke and Quinn describe observables such as
 $A_{0yxz}$, with $x$ and $z$ in the {\it same} direction for
 initial and final states.
 The PS185 collaboration uses for the initial $\bar pp$ state
 the same axes $x,y,z$.
 However, for the final state, they use axes $x'$, $y$ and $z'$
 with $z'$ along the direction of the final $\bar \Lambda$.
 It is necessary to allow for the rotation of spins through
 the scattering angle $\theta$ between initial and final states.
 For triplet states, this involves a simple projection of
 spins as vectors from one set of axes to the other.
 For singlet states, the rotation of axes has no effect.

 \subsection {Parametrisation of Partial Wave Amplitudes}
Partial wave amplitudes need to include three standard factors:
(a) the $1/p$ flux factor for the centre of mass momentum $p$
in the $\bar pp$ channel,
(b) the relativistic phase space factor $\sqrt {\rho _1} =
(2p/\sqrt {s})^{1/2}$ for the $\bar pp$ channel and the factor
$(2k/\sqrt {s})^{1/2}$ for $\bar \Lambda \Lambda $,
(c) Blatt-Weiskopf centrifugal barrier factors for both
$\bar pp$ and $\bar \Lambda \Lambda$ channels [15];
they give the required $k^L$ dependence near threshold on angular
momentum $L$ and momentum $k$ is the $\bar \Lambda \Lambda $ channel.
The product of these three factors will be written as
$G(s)$.
Then partial wave amplitudes for spin $J$, angular momemta
$\ell$ and $L$ in intial and final states
$F_{J,\ell,L}(s)$
are written:
\begin {equation}
F_{J,\ell,L}(s) = G_{J,\ell,L}(s) f_{J,\ell,L}(s),
\end {equation}
where $f(s)$ are analytic funcions.
Note that the factor $G(s)$ must be factored out in order to
avoid branch cuts below threshold.

Data at 1637 MeV/c are adequate to give a unique set of partial
waves.
At other momenta, the analysis reveals quickly that the transition
amplitude $^3S_1 \to \, ^3D_1$ is well determined by the polarisations of
$\Lambda$ and $\bar \Lambda$.
The data are consistent with the same $s$-dependence for this
amplitude as for $^3S_1 \to \, ^3S_1$, except for the centrifugal
barrier factor for the D-wave. The radius of the centrifugal barrier
optimises at $R = 1.1$ fm. To simplify the analysis, the $^3S_1 \to \,
^3S_1$ amplitude is parametrised with coupling constant $g_1$ and the
$^3S_1 \to \, ^3D_1$ transition amplitude is parametrised with coupling
constant $g_1h_1$, where $h_1$ is a complex constant.

The separation between $^3D_1$ and $^3S_1$ initial states
is sensitive only to polarised target data.
Therefore, the $^3D_1 \to \, ^3S_1$ amplitude is parametrised
with coupling constant $g_1h_1'$, and it is necessary to
assume that $h_1'$ does not vary with $s$.
The same is true for the $^3D_1 \to \, ^3D_1$ amplitude which
is fitted with coupling constant $g_1h_1''$ with
$h_1''$ constant.
Physically, the implication is that the branching ratio of
$\bar \Lambda \Lambda$ does not change with mass.
These assumption are of
little consequence at low momenta because the $L = 2$
centrifugal barrier suppresses the amplitude near threshold
for initial D-states.

Partial waves for $2^+$, $3^-$ and $4^+$ are treated in the
same way.
Care is needed even at 1637 MeV/c in handling the amplitude
for $^3F_2 \to \, ^3P_2$.
With present data, the separation of the four $2^+$
amplitudes, abbreviated as $f_{PP}$, $f_{PF}$, $f_{FP}$ and
$f_{FF}$, is the weakest link in the entire analysis.
The $f_{FF}$ amplitude is small and not a matter for
concern.
The $f_{PF}$ amplitude is well determined by polarisations
of $\Lambda$ and $\bar \Lambda $ and differential cross
sections.
However, the $f_{FP}$ amplitude shows some
tendency to drift upwards in magnitude with only a small
change in $\chi ^2$.
The problem is cured by including into $\chi ^2$ a weak
penalty function which limits its magnitude.
The penalty function adds to $\chi ^2$ a term
$$ \Delta \chi ^2 = \frac {|f_{FP}|^2}{\Delta ^2_{FP}}, $$
and the denominator $\Delta ^2_{FP}$
is adjusted so that this term contributes 9 to $\chi ^2$.
This technique allows $F_{FP}$ to grow if the
data really demands it, but constrains it from running
away with little change in $\chi ^2$.
Below 1637 MeV/c, the fit is insensitive to this restriction,
but above 1637 MeV/c, there may be some sensitivity.
Further data from a polarised target at high  momentum would
solve this possible problem.

For $J^P = 3^-$ and $4^+$, contributions from
$^3G_3 \to \, ^3G_3$, $^3D_3 \to \, ^3G_3$ and $^3F_4 \to $
\newline
$^3H_4$ are
negligible because of centrifugal barriers in $\bar \Lambda \Lambda$.
Both the inverse amplitudes $^3G_3(\bar pp) \to
\, ^3D_3 (\bar \Lambda \Lambda )$ and $^3H_4 \to \, ^3F_4$
are definitely required. Surprisingly, the $^3G_4 \to \, ^3G_4$
is also definitely required; $5^-$ amplitudes are negligible.

\begin{table}[t]
\begin {center}
\begin{tabular}{cc}
\hline
Amplitude & Change in $\chi ^2$ \\\hline
$^3S_1 \to \, ^3S_1$ & 1894 \\
$^3S_1 \to \, ^3D_1$ & 271 \\
$^3D_1 \to \, ^3S_1$ & 56 \\
$^3D_1 \to \, ^3D_1$ & 50 \\
$^3P_0$           & 98 \\
$^3P_1$           & 248 \\
$^3P_2 \to \, ^3P_2$ & 1337 \\
$^3P_2 \to \, ^3F_2$ & 69 \\
$^3F_2 \to \, ^3P_2$ & 749 \\
$^3F_2 \to \, ^3F_2$ & 34 \\
$^3D_2$           & 51 \\
$^3D_3 \to \, ^3D_3$ & 681 \\
$^3G_3 \to \, ^3D_3$ & 177 \\
$^3F_3$           & 242 \\
$^3F_4 \to \, ^3F_4$ & 684 \\
$^3H_4 \to \, ^3F_4$ & 111 \\
$^3G_4$           & 108 \\\hline
$^1S_0$           & 15 \\
$^1P_1$           & 25 \\\hline
\end{tabular}
\caption{Changes in $\chi ^2$ when individual partial waves
are dropped from the fit and other amplitudes are re-optimised.}
\end {center}
\end{table}

Table 1 shows changes in $\chi ^2$ when partial waves are
removed from the final fit one by one and remaining
amplitudes are re-optimised.
The singlet partial waves $^1S_0$ and $^1P_1$ are very small,
as the PS185 collaboration found earlier.
Any partial waves affecting $\chi ^2$ by $<10$ are eliminated.

The initial fits take $f_J(s)$ to be constants (where possible)
or linear with $s$, except for the threshold cusp.
In no case does the phase decrease with $s$.
In several partial waves, large linear terms were required,
producing phase variations of order $90^\circ$.
An empirical linear fit to the phase begs the question where
the phase originates.
It rapidly became apparent that better fits could be
obtained by allowing a resonant phase variation in some
partial waves.

A resonance with a large width gives an essentially linear
phase variation.
Therefore the final analysis uses constant amplitudes plus
a resonant form for all partial waves, though allowing
the resonance width to go to a large value if the data
prefer the linear phase variation.
This allows a rather flexible parametrisation of the
$s$-dependence.

In the final fit, the cusp in the $^3S_1$ amplitude is
fitted by taking the amplitude
\begin {eqnarray}
f &=& \frac {1}{M^2 - s - m(s) - iM[\Gamma _0 + \Gamma _{\bar \Lambda
\Lambda } (s)]}, \\
\Gamma _{\bar \Lambda \Lambda } &=& C\sqrt {1 - 4M^2_\Lambda /s},
\end {eqnarray}
and taking $m(s)$ from eqn. (6).
The
magnitude of the constant C in the $\Lambda \Lambda $ width is adjusted
so as to reproduce the observed total intensity of the $^3S_1$ and
$^3D_1$ partial waves, and using the same C for coupling to $\bar pp$
and $\Lambda \Lambda$.

\section {Results}
The $\chi ^2$ of the fit is 1377 for 1201 data points and
61 fitted parameters.
This is a similar quality of fit to partial wave analyses of
$NN$  elastic scattering data.

A technical detail is that normalisations of each
set of differential cross sections and integrated cross
sections are varied in accordance with their published
normalisations. This smooths out some scatter amongst the
points, but has negligible effect on fitted amplitudes.
It turns out to be unnecessary to allow normalisations
of polarisation data to adjust in this way.

\begin{figure}
\begin{center}
\epsfig{file=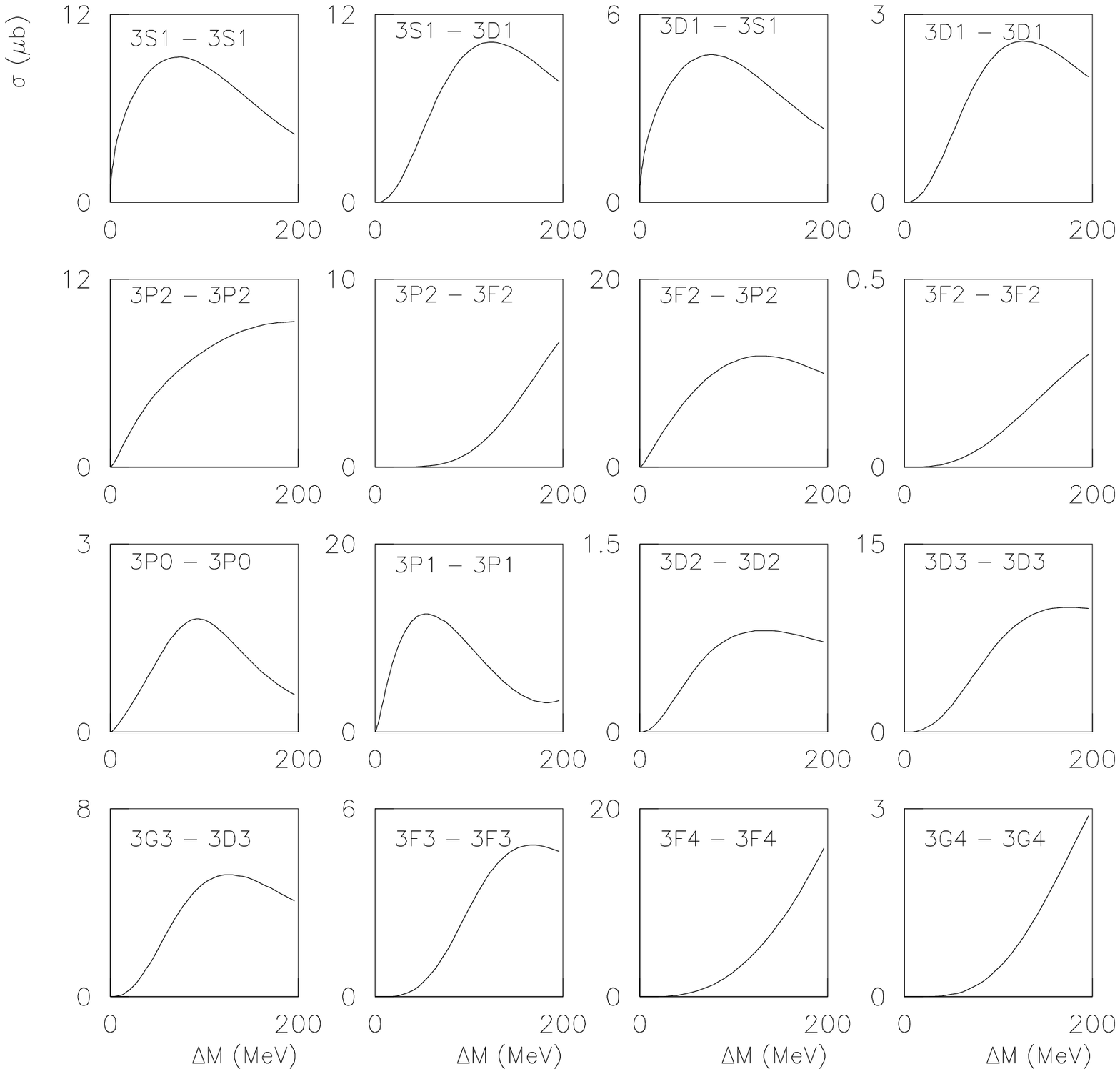,width=15cm}
~\
\caption{ Contributions of partial wave amplitudes to the integrated
cross section;
a beam momentum of 1637 MeV/c corresponds to $\Delta M = 71$ MeV,
$M = 2302.5$ MeV.}
\end{center}
\end{figure}

Fig. 10 shows the intensities of each partial wave in the
integrated cross section.
They are plotted against the mass above the $\bar \Lambda
\Lambda $ threshold: $\Delta M = M - 2M_\Lambda$.
These intensities
contain $G^2(s)$, i.e. the flux and phase space factors and
centrifugal barriers.

It is more instructive to view $|f_J(s)|^2$, where the
kinematic factor $G^2$ is omitted.
These are shown in Fig. 11.
One further factor is also removed.
Each amplitude has Clebsch-Gordan coefficients which affect
the contributions to integrated cross sections.
These factors are listed in Table 2 and are also factored out
in drawing Fig. 11.
The results shows matrix elements squared, unencumbered by
kinematic factors or spin-coupling factors.

\begin{figure} [htb]
\begin{center}
\epsfig{file=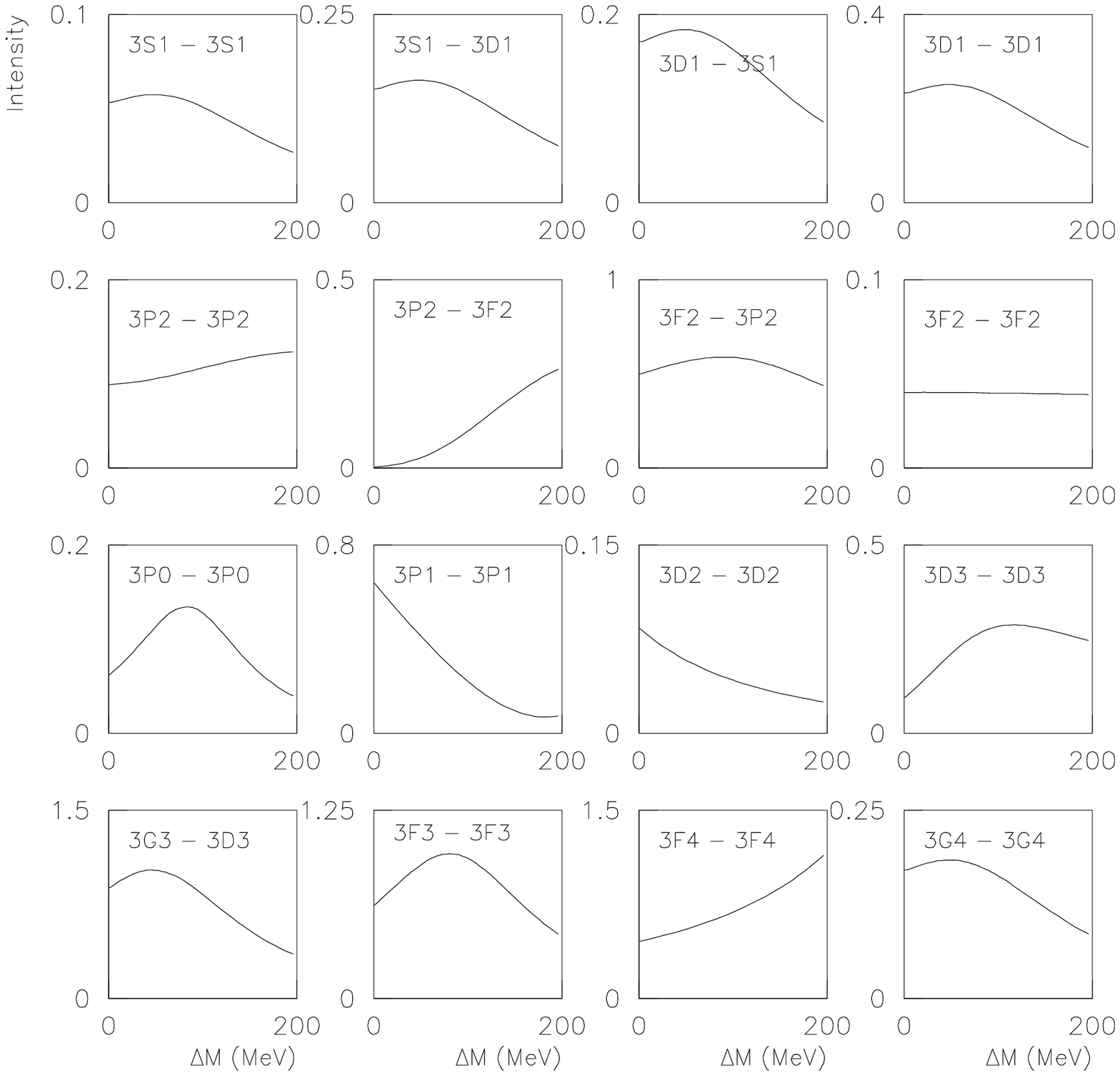,width=15cm}
~\
\caption{ Magnitudes of $|f_J(s)|^2$, i.e. with the kinematic factor
$G^2(s)$ removed;
a beam momentum of 1637 MeV/c corresponds to $\Delta M = 71$ MeV,
$M = 2302.5$ MeV.}
\end{center}
\end{figure}

\begin{table}[htb]
\begin {center}
\begin{tabular}{cc}
\hline
Amplitudes & spin factors \\\hline
$^3S_1$, $^3S_1 \to \, ^3D_1$ & 3 \\
$^3D_1$, $^3D_1 \to \, ^3S_1$ & 3/5 \\
$^3P_0$                    & 1/3 \\
$^3P_1$, $^3D_2$, $^3F_3$, $^3G_4$  & 1  \\
$^3P_2$, $^3P_2 \to \, ^3F_2$ & 5/3 \\
$^3F_2$, $^3F_2 \to \, ^3P_2$ & 5/7 \\
$^3D_3$                    & 7/5 \\
$^3F_4$                    & 9/7 \\
$^3G_3 \to \, ^3D_3$ & 7/9 \\\hline
\end{tabular}
\caption{Spin weighting of amplitudes in integrated cross sections.}
\end {center}
\end{table}

A question is how reliable these intensities are.
A general conclusion is that the final angular
momentum state is well determined by polarisations in
the final state. Hence $^3S_1 \to \, ^3D_1$ and
$^3P_2 \to \, ^3F_2$ intensities are well determined.
In a variety of fits with different combinations of
amplitudes and different assumptions for the fitting
functions $f_J(s)$ and centrifugal barriers,
fluctuations $<10\%$ are observed.
However, identification of the initial state depends
on polarised target data.
Hence the intensities of $^3F_2 \to \, ^3P_2$ and
$^3D_1 \to \, ^3S_1$ partial waves are well determined
($ \pm 7\%)$ at 1637 MeV/c, but their $s$-dependence
away from this mass is uncertain.
Some limitations arise from accurate measurements of
differential cross sections and polarisations, but one
should not draw conclusions from the $s$-dependence of
intensities for $^3F_2 \to \, ^3P_2$ or $^3D_1 \to \, ^3S_1$.
On Figs. 10 and 11, this $s$-dependence is dictated by the centrifugal
barriers.

The top row of Fig. 11 shows $1^{--}$ intensities.
There is a distinct maximum $\sim 60$ MeV above threshold,
i.e. at a mass of 2290 MeV.
It is stronger in $^3D_1 \to \, ^3D_1$ than in $^3S_1 \to \, ^3S_1$.
This peak eventually requires interpretation as a resonance.

The second row of Fig. 11 shows $2^+$ intensities.
The $^3P_2 \to \, ^3P_2$ and $^3F_2 \to \, ^3F_2$ results are
featureless, and the latter is quite small.
However, the $^3P_2 \to \, ^3F_2$ amplitude grows quite
strongly with mass.
It is well determined by polarisations
in the $\bar \Lambda \Lambda$ final state.

The $^3P_0$ and $^3F_3$ intensities show distinct peaks
which will later be associated with known resonances in
Crystal Barrel $I = 0$, $C = +1$ amplitude analyses.
The $^3D_3 \to \, ^3D_3$ and $^3G_3 \to \, ^3D_3$ intensities
likewise show peaks which may be associated with a known
resonance.
The  $^3G_3 \to \, ^3D_3$ amplitude is well determined only
by polarised target data, so the peak in its intensity
follows from the assumption that it scales from the
$^3D_3 \to \, ^3D_3$ amplitude.
The shift between the peaks arises from a mild sensitivity
to differential cross sections at high mass, and may not
be reliable.

The $^3P_1$ and $^3D_2$ amplitudes of Fig, 11 drop from threshold
and cannot be associated with resonant structure.
However, the $^3D_2$ amplitude is small, and it will fit
with very little change in $\chi ^2$ to the known $^3D_2$
resonance $\rho _2(2195)$ [16].
The $^3F_4$ intensity rises steadily with mass and shows
no indication of the known $f_4(2300)$ resonance [17].

The peaks in $^3P_0$, $^3D_3$ and $^3F_3$ fit naturally
as resonances. Fig. 12 shows Argand diagrams. There are
clear loops for these partial waves.
Table 3 shows fitted masses and widths in columns 2 and 3.
Errors cover both statistical variations and systematic
variations over a variety of fits with different assumptions
(e.g. concerning centrifugal barriers and small amplitudes).
In all cases, statistical errors are roughly 35--50\% of
systematic errors.
Around the optimum, both mass and width show well defined
parabolic minima in $\chi ^2$.
For the $1^{--}$ resonance in Table 3, $\Gamma _0 $ of
eqn. (8) is 260 MeV and $\Gamma _{\bar \Lambda \Lambda } = 15$
MeV on resonance, leading to a tabulated width of 275 MeV.

\begin{figure}
\begin{center}
\epsfig{file=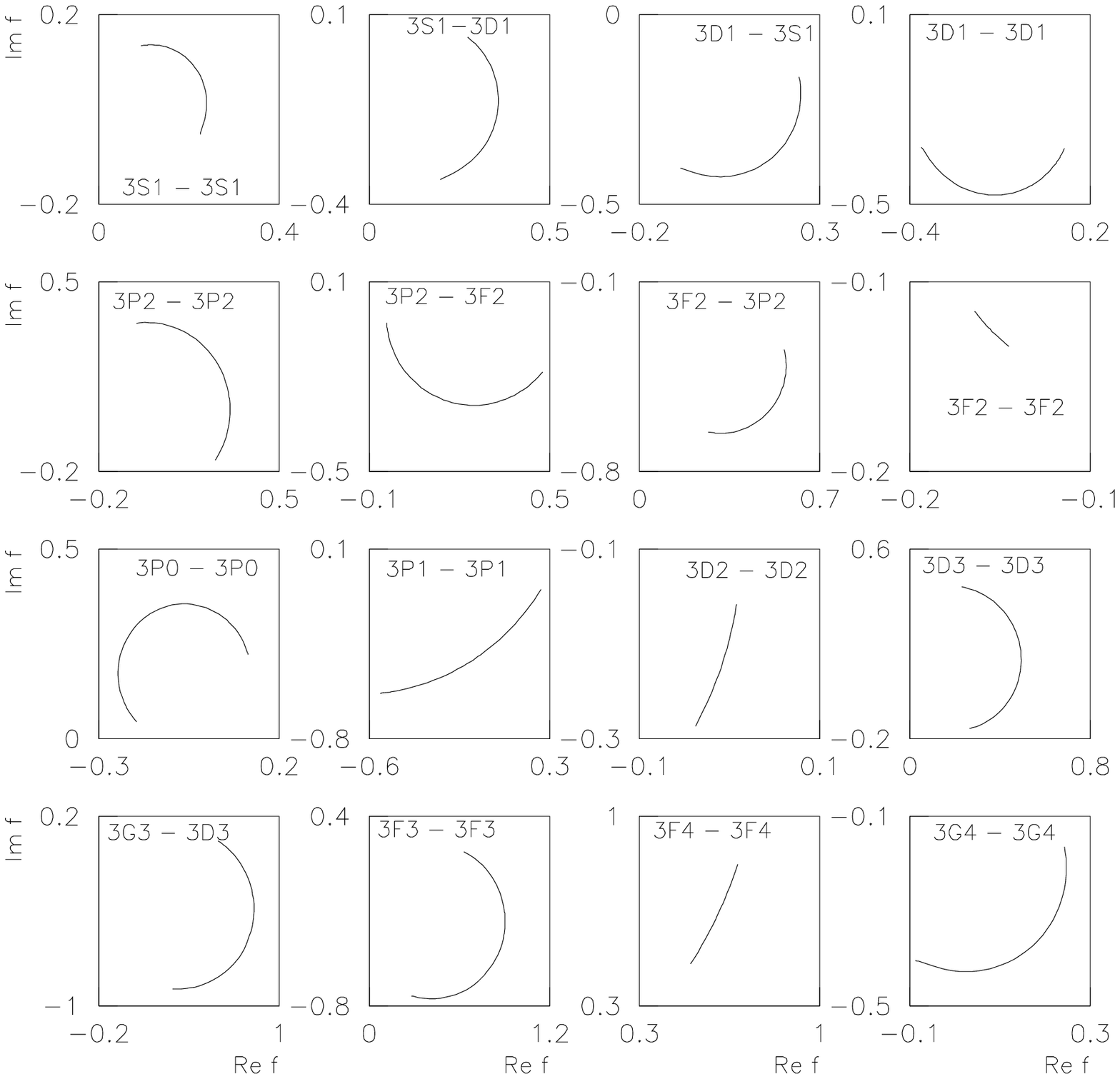,width=15cm}
~\
\caption{Argand diagrams for $f_J(s)$, i.e. after factoring out the
kinematic factor $G(s)$.}
\end{center}
\end{figure}

The next two columns compare with known resonances observed
in Crystal Barrel data [10,16].
Parameters are remarkably close. If the masses and widths
of columns 4 and 5 are used in the fit, the change in
$\chi ^2 $ is only 12, and 6 parameters become fixed.
It therefore looks very likely that the same resonances
appear in PS185 data.

\begin{table}[htb]
\begin {center}
\begin{tabular}{ccccc}
\hline
$J^{PC}$ & $M$(MeV) & $\Gamma $(MeV) & $M$(MeV) & $\Gamma $(MeV)
\\\hline
$0^{++}$ & $2314 \pm 25$ & $144 \pm 20$ & $2337 \pm 14$ & $217
\pm 33$ \\
$2^{++}$ & $2387 \pm 35$ & $33 \pm 100$ &  - & -  \\
$3^{++}$ & $2334 \pm 25$ & $200 \pm 20$ & $2303 \pm 15$ & $214 \pm 29$
\\
$3^{--}$ & $2278 \pm 28$ & $224 \pm 50$ & $2255 \pm 15$ & $175 \pm 30$
\\
$1^{--}$ & $2290 \pm 20$ & $275 \pm 30$ & -  & -  \\ \hline
\end{tabular}
\caption{Columns 2 and 3 show resonance parameters from
PS185 data; columns 4 and 5 show comparisons with Crystal Barrel
results [10,16].}
\end {center}
\end{table}

There is a further feature which agrees with earlier observation
of the $3^-$ resonance.
In Fig. 11, there is a highly significant $^3G_3 \to \, ^3D_3$
intensity. The requirement for this amplitude arises from
$D_{NN}$ and $K_{NN}$ data: dashed curves on Fig. 8 show the
worse fit without this amplitude.
In Ref. [16], a strong $^3G_3$ resonance was likewise
observed at 2255 MeV. In this mass range, both a $^3D_3$ and a $^3G_3$
resonance are expected in conventional quark models. So it is quite
likely that two unresolved $^3D_3$ and $^3G_3$ resonance appear in both
PS185 and Crystal Barrel data. In the Crystal Barrel analysis, a second
state coupling mostly to $^3D_3$ was reported at $2285 \pm 60$ MeV with
$\Gamma = 230 \pm 40$ MeV.

There is also a possible identification of the $2^+$ structure.
There is a known $f_2(1950)$ [17] with a large width of 500
MeV. If it is substituted into the fit, there is almost no change in
$\chi ^2$ and
a small movement downwards of the $2^+$ resonance of Table 3 to 2362
MeV. It is possible that this resonance is to be identified with the
$f_2(2339)$ of Etkin et al. in $\pi \pi \to \phi \phi $ [18];
its appearance in the $\bar \Lambda \Lambda$ channel would not be
surprising. If the $f_2(2339)$ is substituted into the fit with
the width reported by Etkin et al, $\chi ^2$ changes by $< 5$.

The peak in $^3S_1$ and $^3D_1$ at 2290 MeV requires a strong
phase variation.
If  the other peaks described above are identified with known resonances in
Crystal Barrel data, it is inescapable that the $1^{--}$ peak is
resonant. It would be a new resonance. In the Crystal Barrel analysis
of $\omega \eta$, $\omega \pi ^0 \pi ^0$ and $\pi ^- \pi ^+$ channels,
the $1^{--}$ amplitude was not well defined in this mass range. A
resonance at this mass is a natural radial excitation of $\omega _3
(1670)$ [17] and $\omega _3 (1945)$ [16].

\section {Some general remarks}
In earlier work, attempts have been made to fit these
and $\bar pp$ elastic scattering data in terms of meson
exchanges.
There is no conflict between this approach and the apppearance
of resonances.
Meson exchanges can act as part of the driving forces which
generate resonances.
When a resonance appears, the projection of the meson exchange
into an individual partial wave acquires the resonance phase
through re-scattering effects.
A well known example of this is the nucleon exchange term
which partially drives the formation of the $\Delta (1232)$.
Chew and Low showed in 1956 how to include the nucleon exchange
term in an effective range formula which includes the resonance [19].

A little information can be added concerning
$\bar pp \to \bar \Lambda \Sigma ^0$.
Data for the integrated cross section for this process were
reported in Ref. [4] close to threshold.
It is of interest to use detailed balance as in Section 2
to derive the cross section for the inverse process
$\bar \Lambda \Sigma ^0 \to \bar pp$.
Does a cusp appear at threshold?
Results are displayed in Fig. 13.

\begin{figure} [htb]
\begin{center}
\epsfig{file=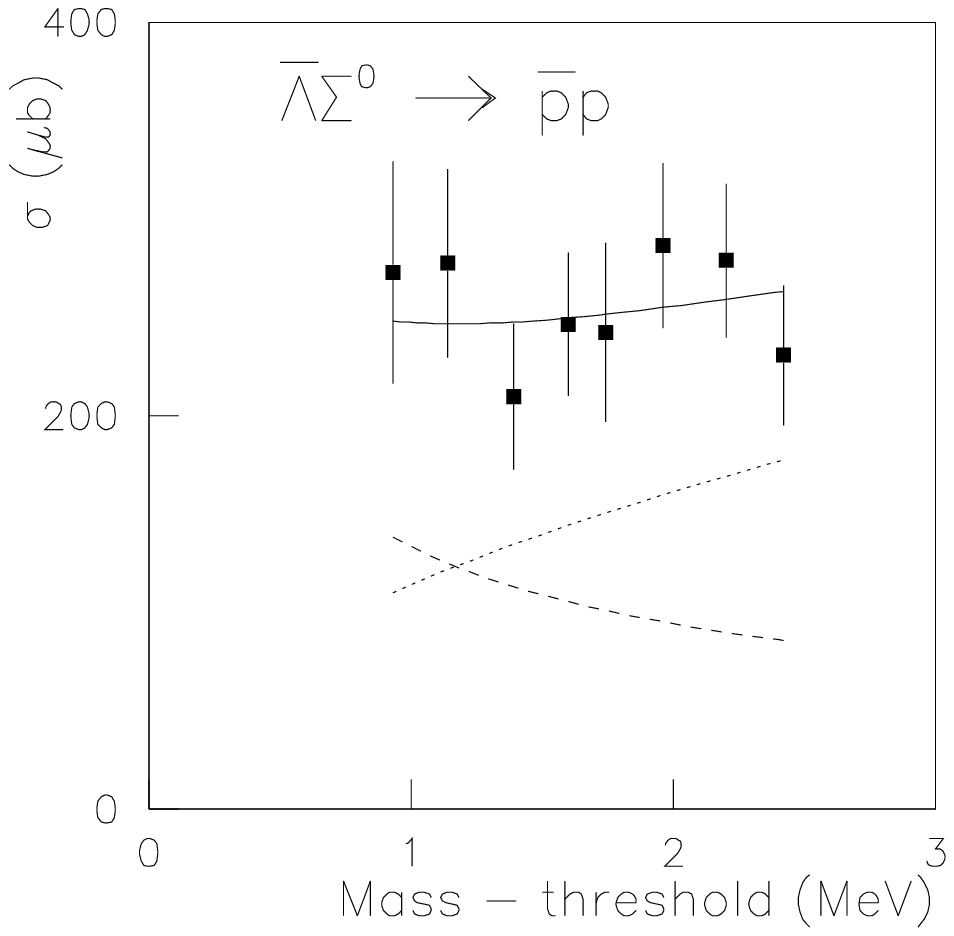,width=6cm} ~\
\caption{Integrated cross sections for $\bar \Lambda \Sigma ^0
\to \bar pp$ deduced from data of Ref. [4]; the full curve
shows a fit to S and P waves, which are shown individually
by the dotted (S) and dashed (P) curves.}
\end{center}
\end{figure}

Errors are sizable, but there is no clear evidence for a
cusp.
The PS185 publication remarks that there is evidence for
strong P-waves very close to threshold.
They are reported to be even stronger than those in
$\bar pp \to \bar \Lambda \Lambda$.
It seems likely that they obscure the presence of a cusp.
A fit is shown using a sum of S and P waves, but there is
considerably flexibility in their relative contributions.

Further progress depends on more data.
It would be valuable to have polarised target data at
other momenta, particularly towards the top of the
mass range, e.g. at 1990 MeV/c.
In principle, such a measurement is feasible at the
new $\bar pp$ ring planned at GSI.
With a frozen spin target, such a measurement is
technically straightforward.
Using a detector such as Crystal Barrel, it would also
be possible to make valuable polarisation measurements
for channels such as $\omega \pi$, $\omega \eta$ and $3\pi ^0$,
allowing a definitive conclusion to the
analysis of Crystal Barrel data.
If a trigger could be included on $K^0_S$ decays and/or $K^0_L$
interactions in the detector,
it would open up the possibility of studying
final states such as $K\bar K$, $K\bar K\pi$ and
$K\bar K \pi \pi$ over a wide mass range, and hence
extending the important LASS data, which run out
around 2100 MeV.

\section {Summary}
A partial wave analysis has been presented of all published
PS185 data.
At 1637 MeV/c, the solution is unique, although care is
needed to restrict the amplitude for $^3F_2 \to \, ^3P_2$
so that it does not drift away to a large value.
The analysis may be extended to cover all other momenta
by making the assumption that the $^3D_1 \to \, ^3S_1$
and $^3D_1 \to \, ^3D_1$
amplitudes are related to  $^3S_1 \to \, ^3S_1$ simply by the
centrifugal barrier for the initial state.
The same assumption is employed for initial $\bar pp$ states
$^3F_2$ and $^3G_3$.
Below 1637 MeV/c, this assumption is not serious, since the
centrifugal barriers for the initial state suppress these
amplitudes strongly.

There is direct evidence for a cusp at threshold in
$\bar \Lambda \Lambda \to \bar pp$.
This cusp needs to be included into the treatment of
the $^3S_1$ partial waves.

There is evidence for large phase variations in several
partial waves in Fig. 12.
If resonances are fitted to $0^{++}$, $3^{++}$ and
$3^{--}$ partial waves, observed resonance parameters
are remarkably close to resonances reported earlier in
Crystal Barrel data.
With that identification, a new $1^{--}$ resonance is
required at 2290 MeV.
Also in $^3P_2 \to \, ^3F_2$, there is evidence for a
resonance around 2360 MeV which fits well as the
$f_2(2339)$ reported by Etkin et al.

\section {Acknowledgments}
I am grateful to Prof. T. Johansson for providing tables of
final data. Also to Dr. K. Paschke and Dr. B. Quinn for
extensive discussions on formulae for observables.

\end{document}